# Ice coverage of dust grains in cold astrophysical environments

Alexey Potapov[1], Cornelia Jäger[1], and Thomas Henning[2]

[1]*Laboratory Astrophysics Group of the Max Planck Institute for Astronomy at the Friedrich Schiller University Jena, Institute of Solid State Physics, Helmholtzweg 3, 07743 Jena, Germany, email: alexey.potapov@uni-jena.de*
[2]*Max Planck Institute for Astronomy, Königstuhl 17, D-69117 Heidelberg, Germany*

**Abstract**

Surface processes on cosmic solids in cold astrophysical environments lead to gas phase depletion and molecular complexity. Most astrophysical models assume that the molecular ice forms a thick multilayer substrate, not interacting with the dust surface. In contrast, we present experimental results demonstrating the importance of the surface for porous grains. We show that cosmic dust grains may be covered by a few monolayers of ice only. This implies that the role of dust surface structure, composition, and reactivity in models describing surface processes in cold interstellar, protostellar, and protoplanetary environments has to be re-evaluated.

# 1. Introduction

Dust grains play a central role in the physics and chemistry of interstellar and circumstellar environments. They influence the thermal properties of the medium by absorption and emission of stellar light, provide surfaces for chemical reactions responsible for the synthesis of a major part of important astronomical molecules from the simplest ones such as $H_2$ and $H_2O$ to complex organic molecules (COMs), e.g. sugars and alcohols, and they are building blocks of comets, asteroids, planetesimals, and planets.

It is well known that dust grains in cold cosmic environments such as molecular clouds and protostellar envelopes and planet-forming disks beyond the snowline are mixed with molecular ices. Typical sketches [1, 2] show a compact dust core surrounded by a thick ice layer. The dust consists mainly of carbon- and silicate-based compounds [3, 4] and the ice includes $H_2O$ (the main constituent accounting for more than 60% of the ice in most lines of sight [5]), CO, $CO_2$, $NH_3$, $CH_4$, $CH_3OH$ and a number of minor components [6, 7]. There is a remarkable diversity in the mixing ratios of molecular ices in interstellar and circumstellar environments. Abundances for some species with respect to $H_2O$ range within more than an order of magnitude. The same is true for cometary ices [8].

Ice column densities are calculated from the band areas of their astronomically observed vibrational transitions using appropriate band strengths measured in the laboratory (see, e.g. [9]). These densities vary dramatically from object to object. The column density of water ice $N(H_2O)$ in prestellar cores and protostellar envelopes determined from the 3 μm band (the cleanest measure of the $H_2O$ ice column density) ranges from $0.3 \times 10^{18}$ to $12 \times 10^{18}$ cm$^{-2}$ [10-12]. Assuming that ice covers a compact dust core and taking a typical value of $10^{15}$ molecules per 1 cm$^2$ per monolayer, we can determine the coverage of grains by water ice to be in the range between 300 and 12000 monolayers (ML) depending on the grain size and the physical conditions (temperature, pressure) of the astrophysical environment. Physical and chemical processes occurring in such ices are barely sensitive to the properties of the dust surface.

On this basis, the majority of the laboratory experiments modelling physical and chemical processes on the surface of cosmic dust grains have been performed on thick multilayer ice mixtures (molecular solids) covering standard laboratory substrates, which might not be characteristic of cosmic dust grains. Adsorption, desorption, and reactivity of different molecules and radicals in molecular solids have been studied extensively during the last decades. We refer the reader to a number of review papers on these topics [1, 13-16].

Compared to molecular solids, there is a handful of studies of the physics and chemistry on the surface of dust grain analogs. We refer to recent papers on the formation [17, 18] and desorption [19] of molecules. Summarising the findings of previous studies, it has been shown that (i) the efficiency of molecule formation depends on the morphology of the grain surface, (ii) the binding energies of species can be quite different for different grain surfaces and on the grain surfaces compared to molecular solids, (iii) functional groups and atoms of the grain surface can participate directly in surface reactions, (iv) the grain surface has a catalytic effect, (v) desorption kinetics and yields of volatile molecules are different for different grain surfaces and compared to molecular solids, (vi) water molecules can be trapped on the grain surface at temperatures above the desorption temperature of water ice.

Thus, dust grains play an important role in processes occurring on their surfaces. If the ice coverage of grains in molecular clouds, protostellar envelopes, and planet-forming disks is much thinner than what is typically assumed, the role of the grain surfaces in the physical and chemical processes in cold astrophysical environments is underestimated and the results obtained in the laboratory studies for molecular solids have to be evaluated in their astrophysical implications.

## 2. The porosity of dust grains

Analysis of cometary dust particles, dust evolution models, and laboratory experiments have indicated that grains in interstellar clouds, protostellar envelopes, and protoplanetary disks may be very porous meaning the existence of a very large surface.

There are a number of mechanisms responsible for the formation of dust grains in interstellar and circumstellar environments. First, nm-sized grains are formed in outflows of evolved stars and then distributed into the interstellar medium (ISM) [3]. Dust extinction observations and models indicate a very efficient growth of dust grains in the ISM (see, e.g. [4] for a review), especially in dense molecular clouds, where grains grow to fractal aggregates of submicron-micron sizes having a porosity of up to 80% [20]. Dust analog materials with the morphological characteristic of interstellar dust can be produced in the laboratory by gas-phase condensation of nm-sized amorphous carbon or silicate grains and their subsequent deposition onto a substrate, where they aggregate (see [21] for a detailed review). The morphology of such grain deposits, both carbon and silicate, can be understood as porous layers of fractal agglomerates having the porosity of up to 90% [22, 23].

In the ISM, dust grains can be completely destroyed by supernova shocks. Estimations showed that only a few percent of the dust produced by stars survive in the ISM [24]. The second mechanism of the dust grains formation is a cold condensation at interstellar conditions. This mechanism was proved in the laboratory and the formation of fluffy, highly porous aggregates similar to the aggregates formed by gas-phase condensation was demonstrated [25, 26].

In protostellar envelopes and planet-forming disks, grains continue to grow. The first stage of dust coagulation in disks leads to the formation of fractal aggregates of submillimetre-millimetre size. Observations of protoplanetary disks suggest that the maximum size of dust grains is 1 mm at least [27]. Laboratory collisional studies (see [28] for a review) and models [29-32] showed that aggregation of μm-sized grain monomers leads to the formation of mm-sized fluffy particles with a fractal dimension of 1 – 2 and the porosity of more than 90%. In protoplanetary disks, such particles are believed to stick to each other due to molecular forces forming kilometre-sized planetesimals that, in turn, form planets due to their gravity. The formation of planets is often accompanied by the formation of debris disks around a central star [33]. In debris disks, a new, top-down mechanism comes into play. Small dust grains can be formed through collisional cascades started by collisions of planetesimals.

The high porosity of interstellar and protoplanetary dust grains is also suggested by the analysis of cometary dust particles. These particles should be the same materials that, mixed with ices, formed comets when the Solar system formed [34]. Analysis of radar [35] and in-situ [36-39] measurements of cometary comae, modelling of the light-scattering properties of the cometary dust [40], and analysis of the samples collected and returned by the Stardust mission [41] showed that cometary dust particles should present a mixture of dense and fluffy, highly porous aggregates. The fluffy particles are typically considered as fractal aggregates of nm-sized grains that may be linked to interstellar dust. Such fractal particles can have microporosity of more than 99% [42].

Thus, there are indications of high porosity of dust grains in interstellar clouds, protostellar envelopes, and planet-forming disks. If this is true and the dust grains in these environments are highly porous particles, they have a corresponding large surface. The surface area of their laboratory analogs cannot be measured precisely but can be estimated.

### 3. The surface area of laboratory analogs of cosmic grains

Our recent laboratory experiments on cosmic dust grain analogs produced by gas-phase condensation and subsequent deposition onto a substrate have demonstrated that the surface area of aggregates of nm-sized grains can be several hundred times larger than the nominal surface area of the dust layer on a substrate, which can be seen as the surface area of an imaginary compact dust core. Our aggregates can be considered as analogs of dust grains in dense molecular clouds, protostellar envelopes, and planet-forming disks, for which they represent building blocks of larger aggregates having the similar structure (due to the fractal growth of grains) and porosity.

In our experiments, the formation of nm-sized amorphous carbon grains is performed by pulsed laser ablation of a graphite target and a subsequent condensation of the evaporated species in a quenching atmosphere of a few mbar of He + $H_2$. The condensed grains are extracted adiabatically from the ablation chamber through a nozzle and a skimmer into a low-pressure chamber generating a particle beam. The beam is directed into the deposition chamber where the grains are deposited onto a KBr substrate. The deposited material is probed in situ by FTIR spectroscopy using a Fourier transform infrared (FTIR) spectrometer (Vertex 80v, Bruker) in the spectral range from 6000 to 400 $cm^{-1}$. For details of the experimental setup see [22, 43]. The thickness of the grain layers is controlled by a quartz crystal resonator microbalance using known values for the deposit area of 1 $cm^2$ and a density of 1.7 g $cm^{-3}$.

In Figure 1, we show electron microscopy images of amorphous carbon grains, where a high porosity and large surface are clearly observable. The left, high-resolution transmission electron microscopy image shows an agglomerate of nanometre-sized primary carbon grains. The grains are composed of small, strongly bent graphene layers or fullerene fragments, which are linked by aliphatic bridges. The right image is a field-emission scanning electron microscopy image. Visible individual grains are still agglomerates of smaller grains. Dust particles on a substrate consist of individual particles in the size range of less than 3-4 nm and large particle aggregates in the size range of up to several tens of nm.

It was shown that water ice, having the nominal thickness of 400 ML or 130 nm (taking an approximate monolayer thickness of 0.3 nm [44]), mixed with carbon grains, thermally desorbs following the first-order kinetics of desorption. In contrast, the same number of water layers desorbs from pure KBr following the zeroth-order kinetics [43]. This result was explained by the desorption of a monolayer of water ice from a large surface of carbon grains. Therefore, the real surface of dust grains can be 400 times larger than the nominal surface of the dust layer. A similar result was obtained in another study, where the monolayer-multilayer transition for $NH_3+CO_2$ ices was observed at the nominal ice thickness of about 200 nm or 600 ML [17].

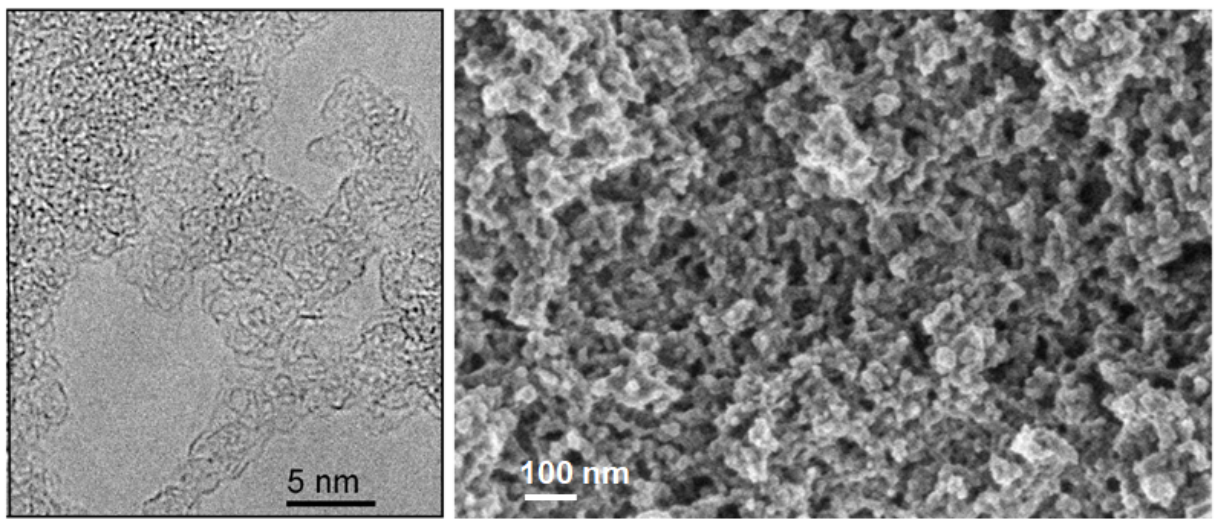

Figure 1. Electron microscopy images of amorphous carbon grains produced by gas-phase condensation and subsequent deposition onto a substrate. The left, high-resolution transmission electron microscopy image shows the porous structure of the condensed carbonaceous grains. The agglomerate of nm-sized primary grains, which are attached to the Lacey carbon support film, is visible in the left upper corner. The right image is a field-emission scanning electron microscopy image of the porous grain layer. Visible individual grains are still agglomerates of smaller grains.

In addition, in the present study, we investigated the temperature-programmed desorption (TPD) of CO ice from the surface of amorphous carbon grains. CO ices of different thicknesses ranging from 0.7 to 310 nominal ML were deposited onto a pure KBr substrate and onto pre-deposited grain layers of 10, 35, and 60 nm thickness at a substrate temperature of 8 K and a pressure in the deposition chamber of $5 \times 10^{-8}$ mbar. The gas phase condensation of grains was performed in a quenching atmosphere of pure He. The ice thicknesses were determined from the 2140 cm$^{-1}$ band area using the band strength of $1.1 \times 10^{-17}$ cm molecule$^{-1}$ (Hudgins, et al. 1993).

TPD experiments were performed by linear ramping of the substrate temperature with a rate of 2 K/min in the temperature range between 8 and 50 K. The error of the temperature measurements was determined to be ±1 K. Infrared spectra during the warming-up were measured with a resolution of 1 cm$^{-1}$ using the FTIR spectrometer. The values for the desorption rate were obtained by taking the first temperature derivative of the integrated intensity of the CO stretching band.

Figure 2 presents the TPD curves for CO ices of different nominal thicknesses deposited onto a pure KBr substrate and onto a 60-nm layer of carbon grains. Zeroth-order multilayer desorption is typically characterized by coinciding leading edges of the desorption curves and a shift of the desorption peak temperature with varying ice thickness. First-order monolayer desorption is typically characterized by no overlap between desorption curves taken for different ice thicknesses and a constant peak temperature.

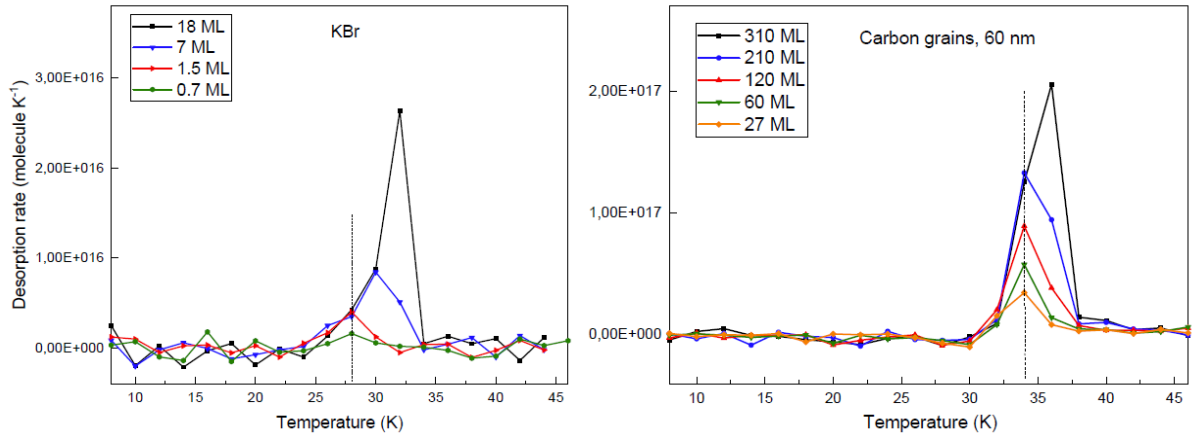

Figure 2. TPD curves for CO ices of different nominal thicknesses deposited onto a pure KBr substrate (left) and onto a 60-nm layer of carbon grains (right). The dashed lines show constant peak temperatures indicating the monolayer ice coverage.

The dashed lines in Figure 2 show constant peak temperatures indicating the monolayer ice coverage. As the reader can see, the monolayer-multilayer transition occurs on KBr at the nominal ice thickness of about 1 ML while on carbon grains the transition is observed at the nominal ice thickness of about 200 ML.

The dependence of the nominal ice thickness corresponding to the monolayer-multilayer transition on the thickness of the grain layer is presented in Figure 3. The growth of the dust layer should lead to an increase of the dust surface area, which depends linearly on the layer thickness. A deviation from the linear dependence should indicate a change in the porosity, which is definitely the case for the substrate change from KBr to carbon grains. For carbon grain layers with the thickness from 10 to 60 nm, the dependence is linear pointing to a constant porosity. The dust growth leads to an increased dust surface area and a corresponding larger number of molecules needed to cover this surface.

The further dust growth taking place in dense clouds, protostellar envelopes, and planet-forming disks, which is hard to mimic in our experimental conditions, can lead to even larger number of molecules needed for one monolayer coverage on the surface of real cosmic grains.

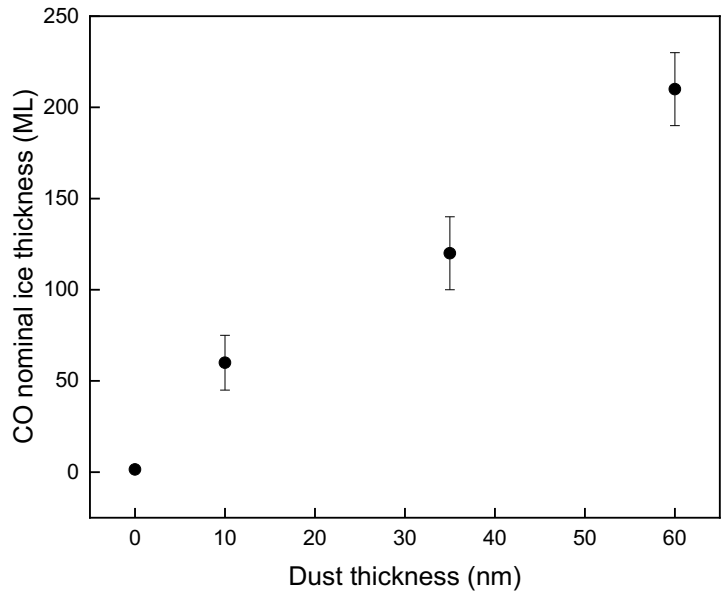

Figure 3. The dependence of the nominal ice thickness corresponding to the monolayer-multilayer transition on the thickness of the grain layer.

## 4. Discussion

Different correction factors for the dust surface area were obtained using TPDs of different molecules: 400 for $H_2O$ [43], 600 for $NH_3+CO_2$ [17], and 200 for CO (this study). The carbon dust thicknesses in these three experiments were: 35 nm for $H_2O$, 70 nm for $NH_3+CO_2$, and 60 nm for CO. It is clear that the correction factor is defined not only by the surface area but also by the surface properties, such as binding energy of molecules, number of available binding sites on the surface, hydrophobic and hydrophilic properties of grains, etc.

Applying the correction factor of 400 to the water ice thickness on cosmic grains of 300 - 12000 ML, we obtain the coverage of grains by water ice in interstellar, protostellar, and protoplanetary environments in the range between 0.75 and 30 ML. Thus, in prestellar, protostellar, and protoplanetary phases, we might have ice layers ranging from a sub-monolayer to a few monolayers instead of a thick multilayer ice coverage on grains. In Figure 4, we offer a new sketch showing cosmic dust grains covered by ice molecules. Of course, such coverage cannot be homogeneous, and we should have clusters, islands, and voids.

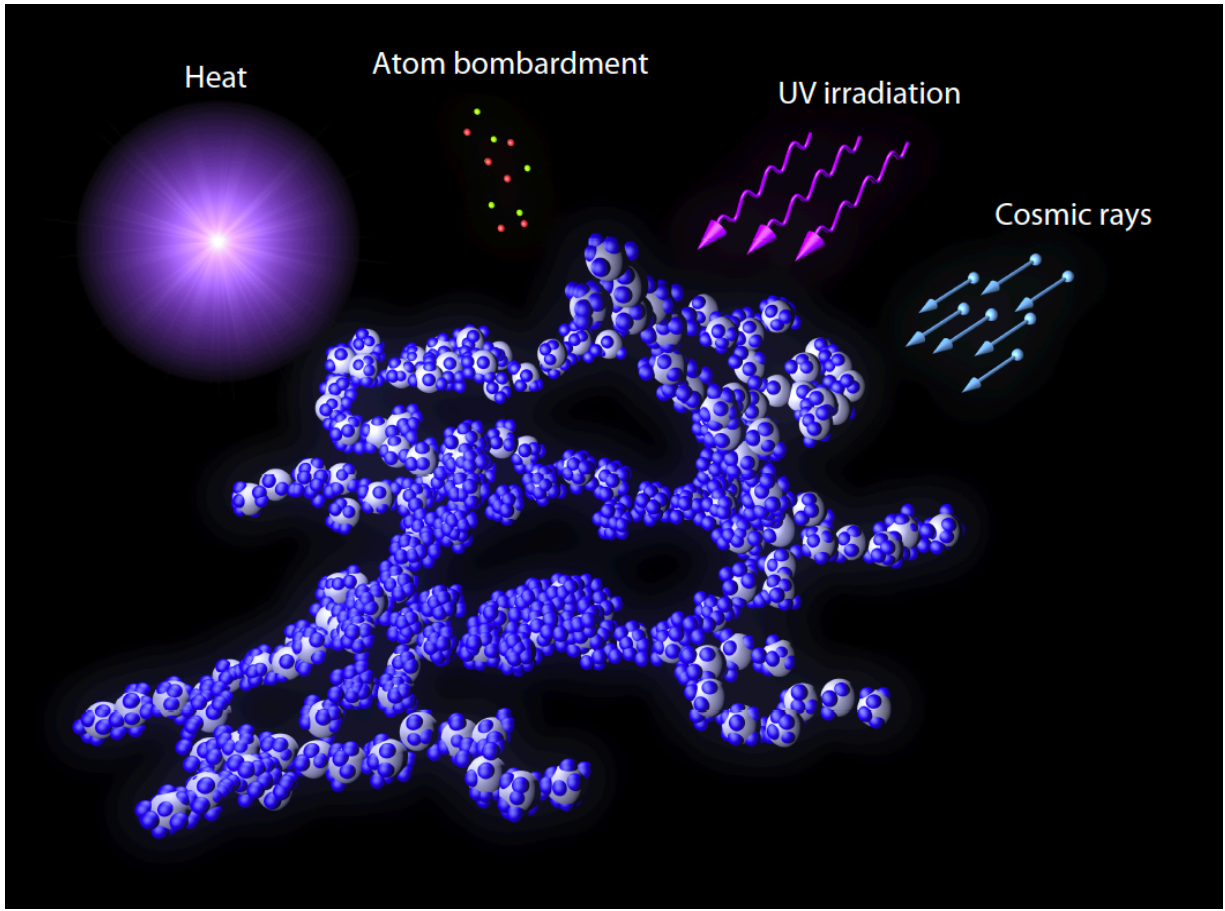

Figure 4. Schematic figure showing dust grains (in grey) mixed with ice molecules (in blue) and the main sources of their processing in astrophysical environments.

Our results reinforce the recent proposition from Marchione et al. [45] that, in contradiction to the popular Onion model, bare dust grain surface in cold astrophysical environments would be available for other species except water to adsorb onto. The authors base their idea on the experimentally demonstrated agglomeration of $H_2O$ molecules on the dust surface [46] that may result in grain surfaces presenting both wet ($H_2O$ presents as agglomerates) and dry (absence of $H_2O$) areas. In our case, the presence of wet and dry areas is simply due to the large surface of grains and its sub-monolayer coverage. A low ice coverage should not prevent agglomeration of $H_2O$ on grains and may even lead to a more efficient agglomeration due to the catalytic role of dust surfaces [17, 47].

If dust grains in cold astrophysical environments are covered by ice layers ranging from sub-monolayers to a few monolayers, experiments such as the formation of CO and $CO_2$ involving atoms of different carbon surfaces covered by $H_2O$ ice by UV, ion, and proton irradiation [48-53], the formation of formaldehyde $H_2CO$ by the addition of O and H atoms to bare carbon grains [18], and the formation of ammonium carbamate $NH_4^+NH_2COO^-$ in the thermal reaction $CO_2$ + $2NH_3$ catalyzed by the surface of carbon and silicate grains [17, 47] become extremely important. They should present more reliable pathways for the formation of molecules in surface reactions in interstellar clouds, protostellar envelopes, and planet-forming disks than the pathways obtained in experiments on molecular solids covering standard laboratory substrates.

Formaldehyde and ammonium carbamate are considered as precursors of prebiotic molecules. The origin of life on Earth is one of the most fascinating questions arising from the studies of our planet and the universe. There are two main hypotheses about the source of the organic compounds that could serve as the basis of life: their formation in the primitive Earth atmosphere [54, 55] and in the ISM with a subsequent delivery to Earth on board of meteorites [56, 57]. The proposed mechanisms for the formation of biomolecules in the ISM include reactions on interstellar grains. In this case, an understanding of the role of dust in surface reactions can be crucial for a simulation of the environmental conditions that allow pathways to the formation of prebiotic molecules in the ISM.

Moreover, the dust surface can influence the desorption as well as the adsorption and mobility of volatile atomic and molecular species on grains. Mobility, depending on the binding energy and distribution of binding sites on the surface, defines a possibility for atoms and molecules to meet and react. Ice condensation and formation and ice desorption are important phase transitions from the gas phase to the solid phase and vice versa in interstellar and circumstellar media. Condensation may go so far that cold phases of the ISM are completely depleted of molecules. Interactions with the dust surface may define the efficiencies of the depletion and of the conversion of adsorbed species to more complex molecules. As a good example in this context, we would mention the depletion of elemental oxygen in the ISM, which is a long-standing problem [58, 59]. As much as a third of the total elemental O budget is unaccounted for in any observed form at the transition between diffuse and dense phases of the ISM and as much as a half is missed in dense phases. One explanation could be that oxygen is trapped with another equally abundant and reactive species, such as carbon and hydrogen, in the solid phase [58, 60].

An understanding of all these astrophysically important processes can only be reached by a detailed surface characterization. The studies of physical-chemical processes on dust grains should be continued with an impact on astrochemical modelling involving dust surfaces.

## Conclusions

Our new experiments on the temperature-programmed desorption of CO ice from the surface of laboratory analogs of cosmic carbon grains confirmed the results of the previous studies. Due to the high porosity of the grains, the surface area is a few hundred times larger than the surface of a KBr substrate, on which the ices are normally deposited. The surface of the substrate can be considered as the surface of the grains seen as a compact dust core. On the basis of our results, we propose that dust grains in cold astrophysical environments may be covered by only up to a few monolayers of molecular ices instead of hundreds of monolayers as it is typically assumed. A low ice coverage makes the role of dust grains in the surface processes much more important compared to the case of the thick ice mantle. With this letter, we would like to point the attention of the astrophysical and astrochemical communities to the necessity of further experimental and theoretical studies of physical-chemical processes on the surface of reliable cosmic dust analogs.

## Acknowledgments

We are grateful to Axel Quetz for his assistance with Figure 4. We thank two anonymous referees for questions, suggestions, and corrections that helped to improve the manuscript. This work was supported by the Research Unit FOR 2285 "Debris Disks in Planetary Systems" of the Deutsche Forschungsgemeinschaft (grant JA 2107/3-2). TH acknowledges support from the European Research Council under the Horizon 2020 Framework Program via the ERC Advanced Grant Origins 83 24 28.

## References

[1] D.J. Burke, W.A. Brown, Ice in space: surface science investigations of the thermal desorption of model interstellar ices on dust grain analogue surfaces, Phys Chem Chem Phys, 12 (2010) 5947-5969.
[2] E. Herbst, Three milieux for interstellar chemistry: gas, dust, and ice, Phys Chem Chem Phys, 16 (2014) 3344-3359.
[3] J. Dorschner, T. Henning, Dust metamorphosis in the galaxy, Astronomy and Astrophysics Review, 6 (1995) 271-333.
[4] B.T. Draine, Interstellar dust grains, Annual Review of Astronomy and Astrophysics, 41 (2003) 241-289.
[5] D.C.B. Whittet, Dust in the Galactic Environment (Bristol: Inst. Phys. Publ.), DOI (2003).
[6] A.C.A. Boogert, P.A. Gerakines, D.C.B. Whittet, Observations of the Icy Universe, Annual Review of Astronomy and Astrophysics, 53 (2015) 541-581.
[7] A.G.G.M. Tielens, The molecular universe, Reviews of Modern Physics, 85 (2013) 1021-1081.
[8] N. Dello Russo, H. Kawakita, R.J. Vervack, H.A. Weaver, Emerging trends and a comet taxonomy based on the volatile chemistry measured in thirty comets with high-resolution infrared spectroscopy between 1997 and 2013, Icarus, 278 (2016) 301-332.

[9] D.M. Hudgins, S.A. Sandford, L.J. Allamandola, A.G.G.M. Tielens, Midinfrared and Far-Infrared Spectroscopy of Ices - Optical-Constants and Integrated Absorbances, Astrophys J Suppl S, 86 (1993) 713-870.
[10] E.L. Gibb, D.C.B. Whittet, A.C.A. Boogert, A.G.G.M. Tielens, Interstellar ice: The Infrared Space Observatory legacy, Astrophys J Suppl S, 151 (2004) 35-73.
[11] A.C.A. Boogert, K.M. Pontoppidan, C. Knez, F. Lahuis, J. Kessler-Silacci, E.F. van Dishoeck, G.A. Blake, J.C. Augereau, S.E. Bisschop, S. Bottinelli, T.Y. Brooke, J. Brown, A. Crapsi, N.J. Evans, H.J. Fraser, V. Geers, T.L. Huard, J.K. Jorgensen, K.I. Oberg, L.E. Allen, P.M. Harvey, D.W. Koerner, L.G. Mundy, D.L. Padgett, A.I. Sargent, K.R. Stapelfeldt, The c2d Spitzer spectroscopic survey of ices around low-mass young stellar objects. I. H2O and the 5-8 mu m bands, Astrophys J, 678 (2008) 985-1004.
[12] A.C.A. Boogert, T.L. Huard, A.M. Cook, J.E. Chiar, C. Knez, L. Decin, G.A. Blake, A.G.G.M. Tielens, E.F. van Dishoeck, Ice and Dust in the Quiescent Medium of Isolated Dense Cores, Astrophys J, 729 (2011).
[13] C.R. Arumainayagam, R.T. Garrod, M.C. Boyer, A.K. Hay, S.T. Bao, J.S. Campbell, J.Q. Wang, C.M. Nowak, M.R. Arumainayagam, P.J. Hodge, Extraterrestrial prebiotic molecules: photochemistry vs. radiation chemistry of interstellar ices, Chem Soc Rev, 48 (2019) 2293-2314.
[14] H. Linnartz, S. Ioppolo, G. Fedoseev, Atom addition reactions in interstellar ice analogues, International Reviews in Physical Chemistry, 34 (2015) 205-237.
[15] K.I. Öberg, Photochemistry and Astrochemistry: Photochemical Pathways to Interstellar Complex Organic Molecules, Chemical Reviews, 116 (2016) 9631-9663.
[16] P. Theule, F. Duvernay, G. Danger, F. Borget, J.B. Bossa, V. Vinogradoff, F. Mispelaer, T. Chiavassa, Thermal reactions in interstellar ice: A step towards molecular complexity in the interstellar medium, Advances in Space Research, 52 (2013) 1567-1579.
[17] A. Potapov, P. Theule, C. Jäger, T. Henning, Evidence of surface catalytic effect on cosmic dust grain analogues: the ammonia and carbon dioxide surface reaction, ApJL, 878 (2019) L20.
[18] A. Potapov, C. Jäger, T. Henning, M. Jonusas, L. Krim, The Formation of Formaldehyde on Interstellar Carbonaceous Grain Analogs by O/H Atom Addition, Astrophys J, 846 (2017) 131.
[19] A. Potapov, C. Jäger, T. Henning, Photodesorption of water ice from dust grains and thermal desorption of cometary ices studied by the INSIDE experiment, Astrophys J, 880 (2019) 12.
[20] V. Ossenkopf, Dust Coagulation in Dense Molecular Clouds - the Formation of Fluffy Aggregates, Astron Astrophys, 280 (1993) 617-646.
[21] C. Jäger, Laboratory Approach to Gas-Phase Condensation of Particles, in Laboratory Astrochemistry, Wiley-VCH, eds. Schlemmer S., Giesen T., Mutschke H., Jäger C. , DOI (2015) 447.
[22] C. Jäger, H. Mutschke, T. Henning, F. Huisken, Spectral Properties of Gas-Phase Condensed Fullerene-Like Carbon Nanoparticles from Far-Ultraviolet to Infrared Wavelengths, Astrophys J, 689 (2008) 249-259.
[23] T. Sabri, L. Gavilan, C. Jager, J.L. Lemaire, G. Vidali, H. Mutschke, T. Henning, Interstellar Silicate Analogs for Grain-Surface Reaction Experiments: Gas-Phase Condensation and Characterization of the Silicate Dust Grains, Astrophys J, 780 (2014).
[24] B.T. Draine, Perspectives on Interstellar Dust Inside and Outside of the Heliosphere, Space Sci Rev, 143 (2009) 333-345.
[25] S.A. Krasnokutski, G. Rouille, C. Jager, F. Huisken, S. Zhukovska, T. Henning, Formation of Silicon Oxide Grains at Low Temperature, Astrophys J, 782 (2014).
[26] G. Rouille, C. Jager, S.A. Krasnokutski, M. Krebsz, T. Henning, Cold condensation of dust in the ISM, Faraday Discuss, 168 (2014) 449-460.
[27] P. D'Alessio, N. Calvet, L. Hartmann, Accretion disks around young objects. III. Grain growth, Astrophys J, 553 (2001) 321-334.
[28] J. Blum, Dust Evolution in Protoplanetary Discs and the Formation of Planetesimals What Have We Learned from Laboratory Experiments?, Space Sci Rev, 214 (2018).
[29] A. Kataoka, H. Tanaka, S. Okuzumi, K. Wada, Fluffy dust forms icy planetesimals by static compression, Astron Astrophys, 557 (2013).

[30] R. Tazaki, H. Tanaka, S. Okuzumi, A. Kataoka, H. Nomura, Light Scattering by Fractal Dust Aggregates. I. Angular Dependence of Scattering, Astrophys J, 823 (2016).
[31] T. Suyama, K. Wada, H. Tanaka, Numerical simulation of density evolution of dust aggregates in protoplanetary disks. I. Head-on collisions, Astrophys J, 684 (2008) 1310-1322.
[32] K. Wada, H. Tanaka, T. Suyama, H. Kimura, T. Yamamoto, Collisional Growth Conditions for Dust Aggregates, Astrophys J, 702 (2009) 1490-1501.
[33] A.V. Krivov, Debris disks: seeing dust, thinking of planetesimals and planets, Res Astron Astrophys, 10 (2010) 383-414.
[34] D. Brownlee, P. Tsou, J. Aleon, C.M.O. Alexander, T. Araki, S. Bajt, G.A. Baratta, R. Bastien, P. Bland, P. Bleuet, J. Borg, J.P. Bradley, A. Brearley, F. Brenker, S. Brennan, J.C. Bridges, N.D. Browning, J.R. Brucato, E. Bullock, M.J. Burchell, H. Busemann, A. Butterworth, M. Chaussidon, A. Cheuvront, M.F. Chi, M.J. Cintala, B.C. Clark, S.J. Clemett, G. Cody, L. Colangeli, G. Cooper, P. Cordier, C. Daghlian, Z.R. Dai, L. D'Hendecourt, Z. Djouadi, G. Dominguez, T. Duxbury, J.P. Dworkin, D.S. Ebel, T.E. Economou, S. Fakra, S.A.J. Fairey, S. Fallon, G. Ferrini, T. Ferroir, H. Fleckenstein, C. Floss, G. Flynn, I.A. Franchi, M. Fries, Z. Gainsforth, J.P. Gallien, M. Genge, M.K. Gilles, P. Gillet, J. Gilmour, D.P. Glavin, M. Gounelle, M.M. Grady, G.A. Graham, P.G. Grant, S.F. Green, F. Grossemy, L. Grossman, J.N. Grossman, Y. Guan, K. Hagiya, R. Harvey, P. Heck, G.F. Herzog, P. Hoppe, F. Horz, J. Huth, I.D. Hutcheon, K. Ignatyev, H. Ishii, M. Ito, D. Jacob, C. Jacobsen, S. Jacobsen, S. Jones, D. Joswiak, A. Jurewicz, A.T. Kearsley, L.P. Keller, H. Khodja, A.L.D. Kilcoyne, J. Kissel, A. Krot, F. Langenhorst, A. Lanzirotti, L. Le, L.A. Leshin, J. Leitner, L. Lemelle, H. Leroux, M.C. Liu, K. Luening, I. Lyon, G. MacPherson, M.A. Marcus, K. Marhas, B. Marty, G. Matrajt, K. McKeegan, A. Meibom, V. Mennella, K. Messenger, S. Messenger, T. Mikouchi, S. Mostefaoui, T. Nakamura, T. Nakano, M. Newville, L.R. Nittler, I. Ohnishi, K. Ohsumi, K. Okudaira, D.A. Papanastassiou, R. Palma, M.E. Palumbo, R.O. Pepin, D. Perkins, M. Perronnet, P. Pianetta, W. Rao, F.J.M. Rietmeijer, F. Robert, D. Rost, A. Rotundi, R. Ryan, S.A. Sandford, C.S. Schwandt, T.H. See, D. Schlutter, J. Sheffield-Parker, A. Simionovici, S. Simon, I. Sitnitsky, C.J. Snead, M.K. Spencer, F.J. Stadermann, A. Steele, T. Stephan, R. Stroud, J. Susini, S.R. Sutton, Y. Suzuki, M. Taheri, S. Taylor, N. Teslich, K. Tomeoka, N. Tomioka, A. Toppani, J.M. Trigo-Rodriguez, D. Troadec, A. Tsuchiyama, A.J. Tuzzolino, T. Tyliszczak, K. Uesugi, M. Velbel, J. Vellenga, E. Vicenzi, L. Vincze, J. Warren, I. Weber, M. Weisberg, A.J. Westphal, S. Wirick, D. Wooden, B. Wopenka, P. Wozniakiewicz, I. Wright, H. Yabuta, H. Yano, E.D. Young, R.N. Zare, T. Zega, K. Ziegler, L. Zimmerman, E. Zinner, M. Zolensky, Comet 81P/Wild 2 under a microscope, Science, 314 (2006) 1711-1716.
[35] J.K. Harmon, S.J. Ostro, L.A.M. Benner, K.D. Rosema, R.F. Jurgens, R. Winkler, D.K. Yeomans, D. Choate, R. Cormier, J.D. Giorgini, D.L. Mitchell, P.W. Chodas, R. Rose, D. Kelley, M.A. Slade, M.L. Thomas, Radar detection of the nucleus and Coma of Comet Hyakutake (C/1996 B2), Science, 278 (1997) 1921-1924.
[36] M. Fulle, V. Della Corte, A. Rotundi, P. Weissman, A. Juhasz, K. Szego, R. Sordini, M. Ferrari, S. Ivanovski, F. Lucarelli, M. Accolla, S. Merouane, V. Zakharov, E.M. Epifani, J.J. Lopez-Moreno, J. Rodriguez, L. Colangeli, P. Palumbo, E. Grun, M. Hilchenbach, E. Bussoletti, F. Esposito, S.F. Green, P.L. Lamy, J.A.M. McDonnell, V. Mennella, A. Molina, R. Morales, F. Moreno, J.L. Ortiz, E. Palomba, R. Rodrigo, J.C. Zarnecki, M. Cosi, F. Giovane, B. Gustafson, M.L. Herranz, J.M. Jeronimo, M.R. Leese, A.C. Lopez-Jimenez, N. Altobelli, Density and Charge of Pristine Fluffy Particles from Comet 67p/Churyumov-Gerasimenko, Astrophys J Lett, 802 (2015).
[37] M. Fulle, A.C. Levasseur-Regourd, N. McBride, E. Hadamcik, In situ dust measurements from within the coma of 1P/Halley: First-order approximation with a dust dynamical model, Astron J, 119 (2000) 1968-1977.
[38] M.S. Bentley, R. Schmied, T. Mannel, K. Torkar, H. Jeszenszky, J. Romstedt, A.C. Levasseur-Regourd, I. Weber, E.K. Jessberger, P. Ehrenfreund, C. Koeberl, O. Havnes, Aggregate dust particles at comet 67P/Churyumov-Gerasimenko, Nature, 537 (2016) 73-75.
[39] T. Mannel, M.S. Bentley, R. Schmied, H. Jeszenszky, A.C. Levasseur-Regourd, J. Romstedt, K. Torkar, Fractal cometary dust - a window into the early Solar system, Mon Not R Astron Soc, 462 (2016) S304-S311.

[40] J. Lasue, A.C. Levasseur-Regourd, Porous irregular aggregates of sub-micron sized grains to reproduce cometary dust light scattering observations, J Quant Spectrosc Ra, 100 (2006) 220-236.
[41] F. Hörz, R. Bastien, J. Borg, J.P. Bradley, J.C. Bridges, D.E. Brownlee, M.J. Burchell, M.F. Chi, M.J. Cintala, Z.R. Dai, Z. Djouadi, G. Dominguez, T.E. Economou, S.A.J. Fairey, C. Floss, I.A. Franchi, G.A. Graham, S.F. Green, P. Heck, P. Hoppe, J. Huth, H. Ishii, A.T. Kearsley, J. Kissel, J. Leitner, H. Leroux, K. Marhas, K. Messenger, C.S. Schwandt, T.H. See, C. Snead, F.J. Stadermann, T. Stephan, R. Stroud, N. Teslich, J.M. Trigo-Rodriguez, A.J. Tuzzolino, D. Troadec, P. Tsou, J. Warren, A. Westphal, P. Wozniakiewicz, I. Wright, E. Zinner, Impact features on Stardust: Implications for comet 81P/Wild 2 dust, Science, 314 (2006) 1716-1719.
[42] M. Fulle, J. Blum, Fractal dust constrains the collisional history of comets, Mon Not R Astron Soc, 469 (2017) S39-S44.
[43] A. Potapov, C. Jäger, T. Henning, Temperature Programmed Desorption of Water Ice from the Surface of Amorphous Carbon and Silicate Grains as Related to Planet-forming Disks, Astrophys J, 865 (2018) 58.
[44] A. Opitz, M. Scherge, S.I.U. Ahmed, J.A. Schaefer, A comparative investigation of thickness measurements of ultra-thin water films by scanning probe techniques, J Appl Phys, 101 (2007).
[45] D. Marchione, A. Rosu-Finsen, S. Taj, J. Lasne, A.G.M. Abdulgalil, J.D. Thrower, V.L. Frankland, M.P. Collings, M.R.S. McCoustra, Surface Science Investigations of Icy Mantle Growth on Interstellar Dust Grains in Cooling Environments, Acs Earth and Space Chemistry, 3 (2019) 1915-1931.
[46] A. Rosu-Finsen, D. Marchione, T.L. Salter, J.W. Stubbing, W.A. Brown, M.R.S. McCoustra, Peeling the astronomical onion, Phys Chem Chem Phys, 18 (2016) 31930-31935.
[47] A. Potapov, C. Jäger, T. Henning, Thermal formation of ammonium carbamate on the surface of laboratory analogs of carbonaceous grains in protostellar envelopes and planet-forming disks, ApJ, DOI (2020, in press).
[48] V. Mennella, M.E. Palumbo, G.A. Baratta, Formation of CO and CO2 molecules by ion irradiation of water ice-covered hydrogenated carbon grains, Astrophys J, 615 (2004) 1073-1080.
[49] V. Mennella, G.A. Baratta, M.E. Palumbo, E.A. Bergin, Synthesis of CO and CO2 molecules by UV irradiation of water ice-covered hydrogenated carbon grains, Astrophys J, 643 (2006) 923-931.
[50] D. Fulvio, A.C. Brieva, S.H. Cuylle, H. Linnartz, C. Jäger, T. Henning, A straightforward method for Vacuum-Ultraviolet flux measurements: The case of the hydrogen discharge lamp and implications for solid-phase actinometry, Applied Physics Letters, 105 (2014) 014105.
[51] U. Raut, D. Fulvio, M.J. Loeffler, R.A. Baragiola, Radiation Synthesis of Carbon Dioxide in Ice-Coated Carbon: Implications for Interstellar Grains and Icy Moons, Astrophys J, 752 (2012) 159.
[52] T. Sabri, G.A. Baratta, C. Jäger, M.E. Palumbo, T. Henning, G. Strazzulla, E. Wendler, A laboratory study of ion-induced erosion of ice-covered carbon grains, Astron Astrophys, 575 (2015) A76.
[53] J. Shi, G.A. Grieves, T.M. Orlando, Vacuum Ultraviolet Photon-Stimulated Oxidation of Buried Ice: Graphite Grain Interfaces, Astrophys J, 804 (2015) 24.
[54] S.L. Miller, H.C. Urey, Organic Compound Synthesis on the Primitive Earth, Science, 130 (1959) 245-251.
[55] S.L. Miller, A Production of Amino Acids under Possible Primitive Earth Conditions, Science, 117 (1953) 528-529.
[56] J. Oro, Comets and Formation of Biochemical Compounds on Primitive Earth, Nature, 190 (1961) 389-&.
[57] B.K.D. Pearce, R.E. Pudritz, D.A. Semenov, T.K. Henning, Origin of the RNA world: The fate of nucleobases in warm little ponds, Proceedings of the National Academy of Sciences of the United States of America, 114 (2017) 11327-11332.
[58] E.B. Jenkins, A Unified Representation of Gas-Phase Element Depletions in the Interstellar Medium, Astrophys J, 700 (2009) 1299-1348.
[59] C.A. Poteet, D.C.B. Whittet, B.T. Draine, THE COMPOSITION OF INTERSTELLAR GRAINS TOWARD zeta OPHIUCHI: CONSTRAINING THE ELEMENTAL BUDGET NEAR THE DIFFUSE-DENSE CLOUD TRANSITION, Astrophys J, 801 (2015).

[60] D.C.B. Whittet, Oxygen Depletion in the Interstellar Medium: Implications for Grain Models and the Distribution of Elemental Oxygen, Astrophys J, 710 (2010) 1009-1016.